\documentclass{article}
\usepackage{amsmath}
\usepackage{amsthm}
\usepackage{amsfonts}
\usepackage{graphicx}
\usepackage{caption}
\usepackage{enumitem}
\usepackage{hyperref}
\hypersetup{
	colorlinks=true,
	linkcolor=blue,
	filecolor=magenta,
	urlcolor=cyan,
	}
\usepackage{float}
\usepackage{tikz-cd}
\newtheorem{theorem}{Theorem}[section]

\begin{document}
\title{Pseudo-Hermiticity, Martingale Processes, and Non-Arbitrage Pricing.}
\author{Will Hicks}
\maketitle
\begin{abstract}
In \cite{Bjork2}, and \cite{Oksendal2}, financial models based on the Wick product, and White Noise formalism are suggested. Although the original purpose is to incorporate integrals with respect to fractional Brownian motion, it is also pointed out in these articles, that this leads naturally to a quantum mechanical interpretation of the financial market. In this article we pursue this idea further, and in particular show how the framework of quantum probability can be used to construct Martingales, without relying on Brownian integrals. We go on to suggest benefits of doing so, and avenues for future work.
\end{abstract}
\section{Introduction:}
The majority of mathematical models, used to capture the dynamics of the financial markets, are based on the underlying concept of Brownian motion. This is in many ways quite natural. If one assumes that the log returns of the price of a financial asset are independent, and (at least in the short term) identically distributed, then the central limit theorem implies that the distribution of returns will converge to a normal distribution. Furthermore, the variance will increase linearly with time. Thus the representation of the random variables, that drive the market price, as Ito integrals, with respect to Brownian motions, seems an obvious first step.\newline
\newline
Also, by basing the random behaviour of the financial market on Brownian motion, one ends up with partial differential equations, such as the Black-Scholes equation, that can be mapped via changes in the coordinate space, to the standard heat equation (for example see \cite{PHL} chapter 4). Crucially, this means that a wide array of analytic, and numerical, methods are available to find solutions.\newline
\newline
However, the limitations of these simple approaches are well known. For example, long term historical analysis of equity index prices often show Hurst parameters of less than $1/2$, indicating negative autocorrelation in the time-series, and from the perspective of risk neutral derivative pricing theory, the presence of arbitrage. See for example \cite{Bjork2}.\newline
\newline
In \cite {Oksendal2}, the authors address this by introducing the concept of a {\em stochastic test function}, $\psi$. Furthermore, the random variable is now given by the {\em generalized stock price}, $S$. The value $S$ can no longer be observed directly. Instead, the observed stock price is determined by taking the inner product with the stochastic test function: $\bar{S}=\langle S|\psi\rangle$. One can take the view that the generalised price $S$ represents the fluctuating core value for a company, and the stochastic test function: $\psi$ represent the market participants. The authors go on to derive a No Arbitrage Theorem for the generalised stock price: $S$, whilst giving examples of what is termed {\em weak arbitrage}, in the observed price: $\bar{S}$.\newline
\newline
In \cite{Oksendal2}, it is noted that the setup is quantum mechanical in nature. The stochastic process $S$, and the test function representing the market participants $\psi$, exist as operators on, and elements in, a Hilbert space. The objective in this article is to demonstrate how this fact can be exploited using the mathematical framework of quantum probability, and to show some benefits of doing so.\newline
\newline
In \cite{Oksendal2}, the objective is to investigate the possibility of constructing a market model using a random process that is not a semimartingale. With this in mind, the authors construct a stochastic process for $S$ using White Noise formalism, and specifically the Wick-Ito-Skorohod integral with respect to a fractional Brownian motion. In this article, we instead wish to investigate how one can construct different Martingale processes using a new approach based on a pseudo-Hermitian infinitesimal generator.\newline
\newline
We first show how the argument works for simple models, before demonstrating the power of the new approach through the example of stochastic discount factors. Finally, we go on to discuss useful insights that can be obtained through a Bohmian interpretation of the analysis presented.
\section{The Quantum Approach to Financial Modelling.}\label{Q_App}
\subsection{Applying the Riesz Representation Theorem:}
In the approach suggested in \cite{Oksendal2}, we start with a Hilbert space vector $|\psi\rangle$, and a linear operator $\langle S|$:
\begin{itemize}
\item $|\psi\rangle$ represents the state of the market (buyers \& sellers) in which the asset is traded. For example it may contain information about how much market participants are willing to pay for an asset.
\item The operator: $\langle S|$ acts on the market state to return a price.
\end{itemize}
Under this interpretation, the observed stock price: $\bar{S}$, for example when a trade occurs, is determined by taking the inner product:
\begin{align*}
\bar{S}=\langle S|\psi\rangle
\end{align*}
In fact, we can put this into a more conventional quantum mechanical setting using the Riesz Representation Theorem (for example see \cite{Hall} Theorem A.52) which we state here for completeness:
\begin{theorem}
If $\varepsilon:\mathcal{H}\rightarrow\mathbb{C}$ is a bounded linear functional on Hilbert space $\mathcal{H}$, then there exists a unique $\chi\in\mathcal{H}$ such that:
\begin{align*}
\varepsilon(\psi)=\langle\chi|\psi\rangle
\end{align*}
\end{theorem}
We assume that the value of any asset is purely determined by the amount that someone is willing to pay for it. Therefore, we assume the value for the asset $S$ is encoded in the state function $|\psi\rangle$. Then the act of price discovery is determined by an operator, which we call $\hat{S}$. Now the generalized stock price functional from \cite{Oksendal2}: $|S\rangle$, is simply the Riesz representation for $\hat{S}$ acting on the market state: $|S\rangle=\hat{S}|\psi\rangle$.\newline
\newline
Depending on the financial interpretation of the model, the basic axioms of quantum probability could then apply as normal:
\begin{enumerate}
\item[1)] The state of the market (including the traded asset price) is represented by a unit vector $\psi\in\mathcal{H}$.
\item[2)] Things that can be observed in the real world are represented by Hermitian operators on the Hilbert space. For example the traded asset price $S$ is represented by the operator $\hat{S}$.
\item[3)] The expected value for an operator $\hat{S}$ is given by:
\begin{align*}
E^{\psi}[S]=\langle\psi|\hat{S}\psi\rangle
\end{align*}
\item[4)] If the asset price for a market, initially in state $\psi$, is measured with the value $S_0$ (for example through a trade execution), then immediately after the measurement, the market will be found in an eigenstate: $\psi_0$, for the operator $\hat{S}$. The following is assumed to apply:
\begin{align}\label{ev_eqn}
\hat{S}\psi_0=S_0\psi_0
\end{align}
\end{enumerate}
For example, if a trade execution price was $S_0$, we could write the market state immediately after the trade as: $\psi_0(y)=e^{i(S_0-y)}$. From \ref{ev_eqn}, we have: $\hat{S}\psi_0=S_0\psi_0$, and if we carry out a second transaction immediately after the first we get an expected price given by:
\begin{align*}
E^{\psi}[S]=\langle\psi_0|\hat{S}\psi_0\rangle=S_0
\end{align*}
Now, given a market state $\psi$, there are (in general) a range of possible outcomes from a trade execution. The observed price $\bar{S}$ now represents the expected outcome from a trade execution, rather than the observed price.\newline
\newline
However, despite this difference, the core principal of the approach outlined in \cite{Oksendal2} is retained. That the price discovery is controlled via the action of a linear functional $\langle S|$, on a Hilbert space vector: $|\psi\rangle$.
\subsection{The Infinitesimal Generator:}
We must now define the time evolution for the system in question. In \cite{Oksendal2}, this is achieved by applying the Wick-Ito-Skorohod integral with respect to a Fractional Brownian motion, to the stock price operator: $\langle S|$.\newline
\newline
For our purposes, we wish to construct Martingales, and so do not require the formalism based on the Wick product. One simple way of constructing Martingales would be to specify a generator: $\mathcal{A}$, so that $e^{\mathcal{A}t}$ forms a semigroup.\newline
\newline
Under the quantum approach, one assumes the time evolution is controlled by a unitary group, rather than a semigroup. Crucially, we assume the time-evolution operator is invertible, and thus the process is invertible. Based on Stones' Theorem (\cite{Hall}) Theorem 10.15, we represent our infinitesimal generator as, where $\hat{H}$ is a densely defined self-adjoint operator:
\begin{align*}
U(t)=e^{i\hat{H}t}
\end{align*}
We can now apply the time evolution to the market state: $\psi$, using the Schr{\"o}dinger equation:
\begin{align}\label{Sch_Eqn}
\frac{\partial\psi}{\partial t}=i\hat{H}\psi
\end{align}
Equation \ref{Sch_Eqn}, forms the fifth basic axiom of the quantum framework.
\subsection{Remarks on Different Quantum Interpretations:}\label{interp}
We now have a quantum framework, defined by a market state function $\psi$, and an operator $\hat{S}$ that represents the price for a tradeable asset. We can calculate the expected price for the traded asset using the relation:
\begin{align*}
E^{\psi}[S]=\langle\psi|\hat{S}\psi\rangle
\end{align*}
However, care must be taken in when interpreting equation \ref{Sch_Eqn}, in a financial context. Specifically, we must take care when deciding which prices we may represent using a quantum operator: $\hat{S}$, and which prices exist at the classical level.
\subsubsection{Pure Quantum Approach:}
In this approach, we let $\hat{S}$ represent any traded price determined by the market $\psi$. For example, $\hat{S}$ could represent a traded stock price, or a listed option on the same stock. This interpretation is analysed in more detail in \cite{Hicks4}. Here we briefly point out the fact that, in this instance, the fully reversible nature of the time-evolution means that there is no external source of randomness or diffusion and we find the model has strange properties. For example a call option on a listed stock price, with strictly positive payout, can have negative time-value.\newline
\newline
We are therefore led to alternative interpretations for the operator $\hat{S}$.
\subsubsection{Quantum vs Classical Securities:}
One alternative, is to interpret a market: $\psi$ as containing information about a traded stock: $S$. The operator $\hat{S}$ could then act by pointwise multiplication on $\psi$:
\begin{align*}
\hat{S}|\psi\rangle=S\psi(S,t)
\end{align*}
In this way, the probability density function for the asset: $S$ is given by $|\psi(S,t)|^2$. Then, when we consider {\em derivatives} whose value is determined by the underlying stock, we assume we are pricing at the classical level. By applying the Spectral Theorem (eg see \cite{Hall} Theorem 10.4, definition 10.5), the value, at time $t<T$ for a payout $f(S_T)$, dependent on the stock observed at $T$, is given by:
\begin{align*}
V_t=E^{\psi}[f(S_T)]=\langle\psi|f(\hat{S})\psi\rangle=\int_{\mathbb{R}}f(y)|\psi(y,T)|^2dy
\end{align*}
This is one potential interpretation of the approach outlined in this article. We give an alternative interpretation in the next section.
\subsubsection{Arrow-Debreu Securities and the Semi-Classical Approach:}
The Arrow-Debreu security is defined as the derivative that pays out \$1, in the event that the traded underlying price is exactly $S_T$ at maturity $T$. Risk neutral pricing theory requires that the derivative value is given (for $t<T$, and Martingale probability measure $Q$) by:
\begin{equation*}
V_t(S_t,t)=\mathbb{E}^{Q}[\delta_{S_T}]=\int_{\mathbb{R}}\delta(y-S_T)dQ(y)=p(S_T,T|S_t,t)
\end{equation*}
Alternatively, if our quantum state starts in the Dirac state: $\delta_{S_t}$, and $\psi$ represents the solution to the Schr{\"o}dinger equation at time $T$, then we have:
\begin{equation*}
p(S_T,T|S_t,t)=\langle\delta_{S_t}|\psi\rangle=\psi(S_t,t)
\end{equation*}
Under this interpretation, we propose that Arrow-Debreu securities are priced within the quantum framework, as described above. In this way, we can interpret $\psi(S_t,t)$ as the probability density function for the traded asset $S$ at time $t$.\newline
\newline
We then assume that all tradeable securities, observable in the real world, exist as classical functions of the underlying Arrow-Debreu securities. That is, we can construct conventional payouts as integrals over the Arrow-Debreu securities:
\begin{align*}
f(S_T)=\int_{\mathbb{R}}\delta(S_T-y)f(y)dy
\end{align*}
\section{Constructing Martingales using Pseudo-Hermitian Infinitesimal Generators:}\label{QApp_2}
In many conventional models of the financial market, Martingales are constructed based on the Martingale representation theorem (eg see\cite{Oksendal}). One makes the assumption that the core random processes driving fluctuation in market prices are Brownian in nature. The parameters that define the drift \& diffusion components can then be calibrated, for example using the prices of exchange traded instruments, or using historical data.\newline
\newline
In addition to extending existing classical modelling to noncommutative probability spaces, part of the objective in doing this is to allow financial principals to be directly encoded in the dynamics of the Martingale process.\newline
\newline
To illustrate how the quantum approach can replace the Martingale representation theorem discussed above, and the use of Brownian motions \& Wiener processes in Mathematical Finance, we illustrate in this section how this can be achieved using the example of the simple Black-Scholes equation.
\subsection{Step 1: Choosing What to Model}
The first step is to decide what we wish to model. For example, we may decide that we wish to model log returns for a listed equity. This seems reasonable, since relative movements in the listed share price seem more fundamental than the absolute amount. In this case, we write the stock price as:
\begin{equation}\label{step1}
S=e^x
\end{equation}
where $x$ represents the random variable we wish to model. In fact, for the majority of liquid listed equities, one can observed forward prices directly in the market, rather than constructing a forward curve using a discount curve, forecasting dividends etc. Therefore, we ignore dividends \& interest rates, and assume that $S$ represents the market price for the forward.
\subsection{Step 2: Choose Dynamics}
In this step, we will encode the dynamics for the random variable $x$ into a Schr{\"o}dinger equation. The form of the Schr{\"o}dinger equation will control what kind of variable we get. For example, fat-tailed, skewed, Markovian etc.\newline
\newline
There are different ways of interpreting the use of the Schr{\"o}dinger equation in this way. Equation \ref{step2} describes the evolution of a wave-function, and whilst real world measurements associated to this wave-function will have random outcomes, the wave-function evolution is deterministic. We discuss different financial interpretations in more detail in section \ref{interp}.\newline
\newline
For the time being, and for the sake of simplicity, we use a straight forward Hamiltonian function:
\begin{align}\label{step2}
i\frac{\partial\psi}{\partial t}=\hat{H}\psi\\
\hat{H}=-\frac{\sigma^2}{2}\frac{\partial^2}{\partial x^2}+C\nonumber
\end{align}
$\psi(x,t)$ represents the wave function for our random variable: $x$. We define the position operator: $X$ in the usual way:
\begin{align*}
\mathbb{E}^{\psi}[X]=\int_{\mathbb{R}}x|\psi(x,t)|^2dx\\
\mathbb{E}^{\psi}[X^2]=\int_{\mathbb{R}}x^2|\psi(x,t)|^2dx\\
\end{align*}
$C$ represents an operator that can be used to enforce the Martingale condition. For example, in this case we show below how we can use a constant potential to ensure $e^x$ is a Martingale. This will be addressed in section \ref{step3}.
\subsection{Step 3: Changing Variables and Pseudo Hermicity}\label{step3}
So far, we have defined a random variable using the Schr{\"o}dinger equation, and obtained a random variable with the correct dynamics. However, we still do not have a partial differential equation with respect to the original market observable: $S=e^x$. We can construct a valid Schr{\"o}dinger equation using the change of variables: $S=e^x$, although the resulting Hamiltonian is generally Pseudo-Hermitian, rather than Hermitian (see \cite{Baaquie}).
\subsubsection{Brief Introduction to Pseudo-Hermiticity:}
The wave-function $\psi$ is defined as belonging to a Hilbert space. For example, we could define: $\psi\in L^2(\mathbb{R})$. For this choice, the Hilbert space inner product is given by:
\begin{equation}\label{ip_original}
\langle\phi|\psi\rangle=\int_{\mathbb{R}}\overline{\phi(y)}\psi(y)dy
\end{equation}
In this case, the model defined by equation \ref{step2}, will conserve probability in the event that:
\begin{equation*}
\hat{H}^{\dagger}=\hat{H}
\end{equation*}
In fact, as detailed in \cite{Mostafazadeh}, we can define a different inner product (and consequently a different Hilbert space), using a linear Hermitian automorphism: $\eta$. We define:
\begin{equation}\label{ip}
\langle\phi|\psi\rangle_{\eta}=\int_{\mathbb{R}}\overline{\phi(y)}(\eta\psi)(y)dy
\end{equation}
Now, our Hamiltonian: $\hat{H}$ will conserve probability, and have  real valued spectrum, in the event that:
\begin{align}\label{pseudo_herm}
\hat{H}^{\dagger}=\eta\hat{H}\eta^{-1}\\
\hat{H}^{\dagger}\eta=\eta\hat{H}\nonumber
\end{align}
Equation \ref{pseudo_herm}, defines a Pseudo-Hermitian Hamiltonian.
\subsubsection{Finding the New Hamiltonian:}\label{newHam}
In our case, when we transfer from one coordinate system to another, we must conserve the inner products. We have:
\begin{equation*}
\frac{dS}{dx}=e^x
\end{equation*}
So, changing variables: $f(S)=\phi(x)$, and $g(S)=\psi(x)$, we get:
\begin{equation*}
\langle\phi|\psi\rangle=\int_{\mathbb{R}}\overline{\phi(x)}\psi(x)dx=\int_{\mathbb{R}}\overline{f(S)}g(S)dS=\int_{\mathbb{R}}\overline{f(e^x)}g(e^x)e^{-x}dx
\end{equation*}
So, by writing: $(\eta\psi)(x)=e^{-x}\psi(x)$, we can translate from a Schr{\"o}dinger equation defined with respect to $x$, to one with respect to $S$. First assume we can find a positive operator square root for $\eta$. Ie, a positive operator $\rho$, such that $\rho^2=\eta$. Then, we write $\hat{K}=\rho^{-1}\hat{H}\rho$. We have (since $\hat{H}$, and $\hat{\rho}$ are assumed to be Hermitian):
\begin{align*}
\hat{K}^{\dagger}\rho^2=\rho\hat{H}\rho^{-1}\rho^2\\
=\rho\hat{H}\rho\\
=\rho^2\hat{K}
\end{align*}
Therefore, if we can find such an operator: $\rho$, then the Hamiltonian we require is given by: $\rho^{-1}\hat{H}\rho$. In our case, we have: $\rho(x)=e^{-x/2}$, and so:
\begin{align}\label{K_Ham}
\hat{K}\psi=e^{x/2}\bigg(-\frac{\sigma^2}{2}\frac{\partial^2(e^{-x/2}\psi)}{\partial x^2}+Ce^{-x/2}\psi\bigg)\nonumber\\
=-\frac{\sigma^2}{2}\frac{\partial^2\psi}{\partial x^2}+\frac{\sigma^2}{2}\frac{\partial\psi}{\partial x}+\bigg(C-\frac{\sigma^2}{8}\bigg)\psi
\end{align}
We must now adjust the operator: $C$ to ensure the traded underlying: $S=e^x$ is a Martingale under the Hamiltonian: $\hat{K}$.
\subsubsection{The Martingale Requirement:}\label{mart_req}
Note that if $\psi$ is a solution to the Schr{\"o}dinger equation defined by $\hat{H}$ (equation \ref{step2}), and if $\rho$ represents pointwise multiplication by an $x$ function (in this case $\rho(x)=e^{-x/2}$), then: $\rho^{-1}\psi$ is a solution to the Schr{\"o}dinger equation defined by Hamiltonian: $\hat{K}$ (equation \ref{step2}).\newline
\newline
We require that $S=e^x$ is a Martingale under the measure defined by: $\eta$, and the Hamiltonian $\hat{K}$ given by \ref{newHam}. Therefore, the Martingale requirement in this case translates to:
\begin{equation*}
E^{\eta}[e^{x}|x_0]=\langle\delta_{x}|e^x|e^{x/2}\psi\rangle_{\eta}=e^{x_0}
\end{equation*}
If we write the solution to \ref{step2} as:
\begin{equation}\label{ClosedFormPsi}
\psi(x,t)=\frac{1}{\sqrt{2\pi\sigma^2t}}exp\bigg(-\frac{x^2}{2\sigma^2t}-Ct\bigg)
\end{equation}
Writing (without loss of generality) $x_0=0$,and utilising \ref{ip} with $\eta\psi=\eta(x)\psi(x,t)$, we get:
\begin{equation*}
E^{\psi}[e^{x}|x_0=0]=\int_{\mathbb{R}}e^ye^{y/2}\psi(y,t)e^{-y}dy=e^{x_0}e^{\sigma^2t/8}e^{-Ct}
\end{equation*}
Therefore in this case we must set $C$ as a constant, with value $\sigma^2/8$. Finally, writing $\hat{K}$ in terms of the new variable $S=e^x$, we find:
\begin{align}\label{Final_BS_Ham}
i\frac{\partial\psi}{\partial t}=\hat{H}_{BS}\psi\\
\hat{H}_{BS}=-\frac{\sigma^2S^2}{2}\frac{\partial^2}{\partial S^2}\nonumber
\end{align}
Equation \ref{Final_BS_Ham} transforms to the standard Black-Scholes equation, using the mapping $it\rightarrow \tau$. This transformation, from real time to complex time, is often referred to as a ``Wick rotation'' (see for example \cite{Folland}, Chapter 7). A similar Wick rotation can be used to transform the standard (Gaussian) Schr{\"o}dinger equation to the heat equation. In the same way that the Wick rotation transforms a Gaussian wave-function to a Gaussian heat diffusion, the Black-Scholes wave-function is transformed to a Black-Scholes diffusion process. Crucially, by carrying out the transformation:
\begin{align*}
e^{i\hat{H}t}\rightarrow e^{\hat{H}t}
\end{align*}
One changes the unitary group generated by the time-evolution operator into a semigroup, where the inverse axiom no longer holds, and each time evolution mapping is no longer invertible.
\subsection{Remarks on the Operator $C$, and Ideas for Further Development of the Method:}\label{future_dev}
In the example studied in this article (the simple Black-Scholes case), the operator $C$ is a constant potential and has the effect of adjusting the drift in order to ensure that  the traded underlying is a Martingale. However, one can use non-constant $C$ to enforce the Martingale condition in models incorporating more exotic effects. In this section we suggest some ideas for potential future development in this direction.
\subsubsection{Stochastic Discount Factors: One Factor Model}
We are representing our traded underlying as $S=e^x$. In this case, $S$ represents the forward contract for a particular maturity. If $S_0$ represents the spot price for an equity, $S_T$ represents the forward price for maturity: $T$, and $r$ represents a funding rate, then non-arbitrage principals (ignoring dividends) require:
\begin{equation*}
S_T=S_0e^{rT}
\end{equation*}
If holding the stock pays a dividend: $D$ at time $t_d<T$, then again this cash flow should be substracted from the price we can achieve at time $T$, in order for the forward price to be arbitrage free, so that:
\begin{equation*}
S_T=S_0e^{rT}-De^{r(T-t_d)}
\end{equation*}
In general, especially for the major equity indices (FTSE100, Euro Stoxx50, S\&P500) as well as the most liquid individual stocks, there are market makers for forward contracts and dividend forecasts are backed out from these. Furthermore, the volatility of the differentials between forward prices at different maturities is of a lower order of magnitude compared to the volatility of the spot price. For this reason, most models of the equity markets, used by practitioners, only incorporate the randomness of $S_0$. Interest rates, and funding costs, are treated as deterministic variables. Dividend cash flows are either treated as fixed, or treated as a fixed percentage of the spot price: $S_0$.\newline
\newline
There are various classical approaches to modelling the randomness of a term structure of discount factors. For example one can model stochastic interest rates using the Heath Jarrow Morton framework (see \cite{HJM}). However, classical models have a number of drawbacks. In general one must compromise between the fact that a continuous discount curve is an infinite dimensional object, and the tractability of methods for calibration and generating solutions.\newline
\newline
In fact, one can use the operator $C$ to incorporate some of the randomness of these factors into the model. Whilst the difficulties mentioned with classical models will also impact quantum models, these do offer different angles from which to approach the issue. For example, we assume:
\begin{itemize}
\item The dividends \& interest rates can be incorporated into a single discount factor. We write this: $S=DFe^x$.
\item This discount factor is driven by the same combination of factors (both specific to the equity in question and general) that drive the spot price. The discount factor is therefore represented using a function: $DF(x)$. So that we now write our forward price: $DF(x)e^x$.
\item The majority of the time the dividends and interest are stable. In other words we have $DF(x)\approx 1$ for small $x$.
\item Without loss of generality we can incorporate the initial spot price ($S_0$ say) into $DF(x)$ so that the starting value of $x$ is $0$.
\item By modelling interest rates \& dividends using the same underlying variable ($x$) as the main market price, the intention is to keep the model simple from the perspective of calibration \& generating solutions, whilst simultaneously incorporating factors such as cuts to dividends after a market crash.
\end{itemize}
Now we require $DF(x)e^x$ to be a Martingale. Consider, for example, that we wish to take account of the possibility of a cut in interest rates or a cut in the dividend yield under extreme market conditions. This would lead to respectively an increase or a decrease in the relevant discount factor. Then we could apply:
\begin{equation*}
DF(x)=\frac{1}{1+\varepsilon x^2}
\end{equation*}
In this instance, positive $\varepsilon$ would lead to a reduction in the discount factor under extreme market conditions, corresponding to a cut in the dividend yield expectations. Alternatively, a negative value for $\varepsilon$ would lead to an increase in the discount factor under extreme market conditions. This would correspond to a cut in interest rates. Figure 1 shows a chart of the discount factor against the driver: $x$.
\begin{figure}
\includegraphics[scale=1]{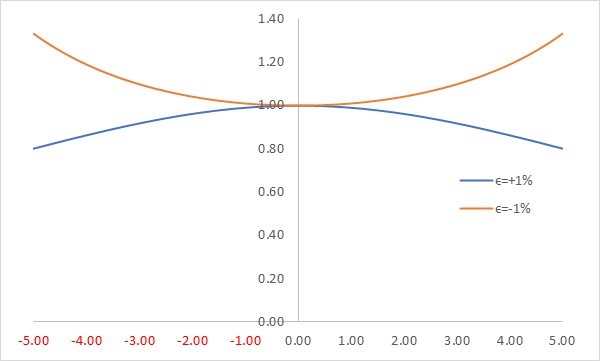}
\caption{Discount factor versus driver for positive \& negative values for $\varepsilon$.}
\end{figure}
Now in order to ensure: $DF(x)e^x$ is a Martingale we require a non-constant potential $C(x)$. Although we can no longer necessarily write out a closed form solution (such as equation \ref{ClosedFormPsi}) we can still apply the Feynman-Kac formula (see for example \cite{Hall} Theorem 20.3) and use Path integral methods to find solutions.
\subsubsection{Stochastic Discount Factors: Multi-Factor Models}
One could relax the above assumption that the random behaviour is driven by a single random variable, by simply extending our Hilbert space from $L^2(\mathbb{R})$ to $L^2(\mathbb{R}^n)$, where $n$ represented the number of random variables. For example, we could model a forward price: $S_T=DF(y)e^x$, whereby our Hilbert space could be: $L^2(\mathbb{R}^2)$. Now, the financial market is represented by wave function: $\psi(\mathbf{x},t)$, where $\mathbf{x}=(x,y)$, and $x$ was the random variable driving the equity markets, and $y$ the random variable driving the interest rates.\newline
\newline
Now, we must use the operator $C$ to ensure $DF(y)e^x$ is a Martingale, and therefore $C$ must again be a non-constant potential: $C(x,y)$. Whilst modelling in a higher number of dimensions may increase the difficulty in finding solutions, the same principals apply. One can still use Path integral methods, based on the Feynman-Kac result.
\section{Hamilton Jacobi Theory and the Bohmian Approach:}\label{B_app}
\subsection{The Bohmian Approach to Step 2:}
In this section, we briefly review the strategy presented in sections \ref{Q_App} and \ref{QApp_2}, from the perspective of Hamilton-Jacobi theory and the Bohmian approach to quantum mechanics. For an overview of this approach, see \cite{Holland}. For an example of an application to quantum finance, see \cite{Ishio_Haven}.\newline
\newline
Start by inserting the wave function:
\begin{equation}\label{HJ_psi}
\psi=Re^{iS}
\end{equation}
Into the Schr{\"o}dinger equation with Hermitian Hamiltonian:
\begin{equation*}
\hat{H}=-\frac{\sigma^2}{2}\hat{P}^2
\end{equation*}
one obtains 2 partial differential equations (see for example \cite{Holland} chapter 3):
\begin{align}
\frac{\partial S}{\partial t}+\frac{\sigma^2(\nabla S)^2}{2}-\frac{\sigma^2}{2}\frac{\nabla^2 R}{R}=0\label{HJE}\\
\frac{\partial^2 R^2}{\partial t^2}+\nabla\big(\sigma^2R^2\nabla S\big)=0\label{CE}
\end{align}
Now, equation \ref{HJE} can be interpreted as a Hamilton-Jacobi equation with ``quantum potential'': $Q=-\frac{\sigma^2}{2}\frac{\nabla^2 R}{R}$. A classical solution $S$ to this equation represents a particle with momentum:
\begin{equation*}
P=\nabla S
\end{equation*}
From a quantum perspective $|\psi|^2=R^2$ represents the probability density function for the particle. Therefore, again from a classical perspective, if we interpret $R^2$ as the density ($\rho$) of a cloud of particles, then the particle flux will be defined by:
\begin{equation*}
\vec{q}=\frac{\sigma^2}{2}\nabla S
\end{equation*}
Finally therefore, we can interpret equation $\ref{CE}$ as the continuity equation for the particle:
\begin{equation}\label{CE_Classical}
\frac{\partial\rho}{\partial t}+\nabla\vec{q}=0
\end{equation}
\subsection{A Pseudo-Hermitian Case: The Black-Scholes Hamiltonian}
We now investigate how this works for the Black-Scholes Hamiltonian. For now, we work using the pseudo-Hermitian Hamiltonian given by equation \ref{K_Ham}. We proceed by inserting the wave function \ref{HJ_psi} into the Schr{\"o}dinger equation defined by the Black-Scholes Hamiltonian.
\begin{align*}
\frac{\partial\psi}{\partial t}=\frac{\partial R}{\partial t}e^{iS}+iR\frac{\partial S}{\partial t}e^{iS}\\
\frac{\partial\psi}{\partial x}=\frac{\partial R}{\partial x}e^{iS}+iR\frac{\partial S}{\partial x}e^{iS}\\
\frac{\partial^2\psi}{\partial x^2}=\frac{\partial^2 R}{\partial x^2}e^{iS}+2i\frac{\partial R}{\partial x}\frac{\partial S}{\partial x}e^{iS}-R\bigg(\frac{\partial S}{\partial x}\bigg)^2e^{iS}
\end{align*}
So, inserting $\psi=Re^{iS}$ into the Schr{\"o}dinger equation:
\begin{align*}
i\frac{\partial\psi}{\partial t}=-\frac{\sigma^2}{2}\frac{\partial^2\psi}{\partial x^2}+\frac{\sigma^2}{2}\frac{\partial\psi}{\partial x}
\end{align*}
We get:
\begin{align*}
i\frac{\partial R}{\partial t}e^{iS}-R\frac{\partial S}{\partial t}e^{iS}=\frac{\sigma^2}{2}e^{iS}\bigg(\frac{\partial R}{\partial x}+R\bigg(\frac{\partial S}{\partial x}\bigg)^2-\frac{\partial^2 R}{\partial x^2}\bigg)+\frac{i\sigma^2}{2}e^{iS}\bigg(R\frac{\partial S}{\partial x}-2\frac{\partial R}{\partial x}\frac{\partial S}{\partial x}\bigg)
\end{align*}
We now divide by $e^{iS}$, and collect together real \& imaginary terms:
\begin{align*}
R\frac{\partial S}{\partial t}+\frac{\sigma^2}{2}\bigg(\frac{\partial S}{\partial x}\bigg)^2+\frac{\sigma^2}{2}\frac{\partial R}{\partial x}-\frac{\sigma^2}{2}\frac{\partial^2 R}{\partial x^2}=0\\
i\frac{\partial R}{\partial t}+i\sigma^2\frac{\partial R}{\partial x}\frac{\partial S}{\partial x}-i\frac{\sigma^2}{2}R\frac{\partial S}{\partial x}=0
\end{align*}
Finally, we divide the real equation by $R$, multiply the imaginary terms by $R$, and replace $\partial /\partial x$ with the more general $\nabla$ to get:
\begin{align}
\frac{\partial S}{\partial t}+\frac{\sigma^2(\nabla S)^2}{2}+\frac{\sigma^2}{2}\frac{\nabla R}{R}-\frac{\sigma^2}{2}\frac{\nabla^2 R}{R}=0\label{HJE2}\\
\frac{\partial R^2}{\partial t}+\nabla\bigg(\frac{\sigma^2R^2\nabla S}{2}\bigg)-\sigma^2R^2\nabla S=0\label{CE2}
\end{align}
Now, equation \ref{HJE2} can still be interpreted as a Hamilton-Jacobi equation, with new quantum potential: $Q=\frac{\sigma^2}{2}\frac{\nabla R}{R}-\frac{\sigma^2}{2}\frac{\nabla^2 R}{R}$. However, the interpretation of equation \ref{CE2} as a continuity equation no longer works in the same way. This is a consequence of the fact that the pseudo-Hermitian Hamiltonian given by equation \ref{K_Ham}, does not conserve probability in the standard Hilbert space inner product.\newline
\newline
However, we know from \cite{Jana} and \cite{Mostafazadeh}, that the Hamiltonian conserves probabilities under the inner product \ref{ip}, with $\eta=e^{-x}$. Therefore, the probability density is now given by: $\rho=e^{-x} R^2$. Inserting this into the classical continuity equation \ref{CE_Classical} we get:
\begin{equation}\label{CE_interim}
\frac{e^{-x}\partial R^2}{\partial t}+\nabla\big(e^{-x}\sigma^2R^2\nabla S\big)=0
\end{equation}
After multiplying through by $e^x$ we get back to equation \ref{CE2}.
\subsection{Interpretation of of the Quantum Potential:}
To start with, consider the Hamilton-Jacobi equation for a particle, obeying classical mechanics with zero potential (see \cite{Holland}). The partial differential equation is given by:
\begin{equation}\label{free_part}
\frac{\partial S}{\partial t}+\frac{\sigma^2(\nabla S)^2}{2}
\end{equation}
The function $S$, which is itself determined by the relevant initial conditions, determines the motion of the free particle through the relationship:
\begin{equation*}
P=-\nabla S
\end{equation*}
In the absence of any potential, the particle simply moves at constant velocity. If we introduce a large number of particles, each with different initial conditions, then each particle will move with a different velocity, depending on the initial condition. In the absence of an external potential, these velocities will be constant.\newline
\newline
Therefore, consider the situation of $N$ such particles moving freely along the real number line, having started at $x=0$. The position of the particle: $i$, with velocity: $v_i$, after time: $t$, will be given by: $X_i=v_it$. Therefore the variance of the particles after time: $t$ is given by:
\begin{align}
\mathbb{E}[X^2]=\frac{1}{N}\sum_{i=1}^N (v_it)^2\nonumber\\
\mathbb{E}[X]=\frac{1}{N}\sum_{i=1}^N v_it\nonumber\\
\mathbb{E}[X^2]-\mathbb{E}[X]^2=\frac{t^2}{N}\bigg(\sum_{i=1}^N (v_i)^2-\Big(\sum_{i=1}^N v_i\Big)^2\bigg)
\end{align}
We note the following:
\begin{itemize}
\item The variance increases with $t^2$. Therefore, this cannot represent a diffusion with independent increments.
\item In fact, each time step for one of the particles is 100\% correlated, with the previous time-step. This is natural, since each particle is moving freely under constant velocity.
\item If we wish the $N$ particles to represent particles moving under a random diffusion, such as Brownian motion, then we must introduce a potential function.
\end{itemize}
Thus, if we interpret the wave-function as defining the probability distribution for a classical diffusion, then the quantum potential $Q$ defines the potential function that will ensure the classical particles have the correct statistical properties.\newline
\newline
Put another way, consider $N$ particles moving under quantum potential: $Q$. These particles, whilst obeying the laws of classical mechanics, and depending on the initial conditions, would have a probability distribution consistent with the Schr{\"o}dinger wave-function. In some senses one could interpret the randomness as arising simply from the variation in the initial conditions. Once the particles were set in motion, the system is deterministic.\newline
\newline
In fact, the setting of the initial conditions, can be defined by the moments one wishes the probability density function to have. This is the equivalent to specifying the mean and variance of the diffusion process:
\begin{align*}
\lim_{N\rightarrow\infty} \sum_{i=1}^N v_i^2\rightarrow \sigma^2\\
\lim_{N\rightarrow\infty} \sum_{i=1}^N v_i\rightarrow \mu
\end{align*}
\subsubsection{Brownian Example:}
The fundamental solution to equation \ref{step2} with $C=0$, is given by:
\begin{align}\label{psi_R}
\psi(x,t)=K_t(x)*\widetilde{\psi_0}(p)\nonumber\\
K_t(x)=\frac{1}{\sqrt{2\pi\sigma^2it}}exp\Big(\frac{ix^2}{2\sigma^2t}\Big)
\end{align}
Where $f*g$ represents the convolution: $\int f(x-y)g(y)dy$, and $\widetilde{h}(p)$ the Fourier transform of $h(x)$. We note, that under the Wick rotation, $\tau=it$ we get back to a heat kernel:
\begin{equation*}
K_{\tau}(x)=\frac{1}{\sqrt{2\pi\sigma^2\tau}}exp\Big(-\frac{x^2}{2\sigma^2\tau}\Big)
\end{equation*}
Following the analysis in \cite{Holland} chapter 3, we set:
\begin{align*}
R=(\overline{\psi}\psi)^{1/2}\\
=\frac{1}{2\pi\sigma^2\tau}exp\Big(\frac{x^2}{2\sigma^2\tau}\Big)
\end{align*}
Finally, we can calculate the quantum potential:
\begin{align}
Q=-\frac{\sigma^2}{2}\frac{\nabla^2 R}{R}\nonumber\\
=\frac{\sigma^2}{2}\bigg(\Big(\frac{x}{\sigma^2t}\Big)^2-\frac{1}{\sigma^2t}\bigg)
\end{align}
So finally, the Quantum potential in this case acts as a quadratic potential, that becomes shallower over time.
\subsubsection{Choosing the Quantum Potential:}
Therefore, it is clear from the analysis above, that one potential future avenue of research is to design diffusion processes, to be applied to finance, by starting with the base Hamilton-Jacobi equation for a free particle. The simplest case being equation \ref{free_part}, before choosing the quantum potential: $Q$.\newline
\newline
We have shown that using a quadratic potential well for the quantum potential results in a a Black-Scholes model. Applying other quantum potentials, which can be tailored to the desired dynamics, will result in different market models. In fact a variety of desirable effects, including different values for the Hurst parameter, and fat-tails, can be introduced in this way.

\end{document}